\documentclass[pre,onecolumn,notitlepage,longbibliography,amsmath,amssymb,floats,superscriptaddress,nofootinbib,10pt]{revtex4-2}
\usepackage{graphicx,color}
\usepackage{amsmath}
\usepackage{amsfonts, amsbsy}
\usepackage{amssymb}
\usepackage{eqnarray}
\usepackage{bm}
\usepackage{blkarray}
\usepackage{mathtools}
\usepackage{psfrag}
\usepackage{multirow}
\usepackage{textcomp}
\usepackage{siunitx}
\usepackage{lipsum}
\usepackage[title]{appendix}
\usepackage{soul}
\usepackage{enumitem}
\usepackage{soul}
\usepackage[sort&compress]{natbib}
\usepackage[breaklinks,colorlinks = true,linkcolor = blue,urlcolor  = blue,citecolor = blue,anchorcolor = blue,hyperindex=true, linktocpage=true]{hyperref}
\usepackage{float}

\usepackage[dvipsnames]{xcolor}

\definecolor{nblue}{RGB}{28,130,185}

\newcommand{\be}{\begin{equation}}
\newcommand{\ee}{\end{equation}}

\newcommand{\bg}{\begin{gathered}}
\newcommand{\eg}{\end{gathered}}

\begin{document}
\title{Lift force in chiral, compressible granular matter}

\author{Jaros\l{}aw Paw\l{}owski}
\affiliation{Institute of Theoretical Physics, Wroc\l{}aw University of Science and Technology, 50-370 Wroc\l{}aw, Poland}

\author{Marcin Dudziak}
\affiliation{Institute of Theoretical Physics, Wroc\l{}aw University of Science and Technology, 50-370 Wroc\l{}aw, Poland}

\author{Matteo Baggioli}
\affiliation{School of Physics and Astronomy, Shanghai Jiao Tong University, Shanghai 200240, China}
\affiliation{Wilczek Quantum Center, School of Physics and Astronomy, Shanghai Jiao Tong University, Shanghai 200240, China}
\affiliation{Shanghai Research Center for Quantum Sciences, Shanghai 201315, China}

\author{Jie Zhang}
\affiliation{School of Physics and Astronomy, Shanghai Jiao Tong University, Shanghai 200240, China}
\affiliation{Institute of Natural Sciences, Shanghai Jiao Tong University, Shanghai 200240, China}

\author{Piotr~Sur\'owka}
\affiliation{Institute of Theoretical Physics, Wroc\l{}aw University of Science and Technology, 50-370 Wroc\l{}aw, Poland}
\affiliation{Institute of Condensed Matter Physics, Department of Physics, Technical University of Darmstadt, Hochschulstr. 8,
64289, Darmstadt, Germany}

\begin{abstract}
Micropolar fluid theory, an extension of classical Newtonian fluid dynamics, incorporates angular velocities and rotational inertias and has long been a foundational framework for describing granular flows. We propose a macroscopic model of granular matter based on micropolar fluid dynamics, which incorporates internal rotations, couple stresses, and broken parity through odd viscosity. Our framework extends traditional micropolar theory to describe chiral granular flows driven far from equilibrium, where energy is continuously injected and dissipated. In particular, we focus on steady states and explicitly neglect energy conservation, reflecting the dissipative nature of granular systems maintained in non-equilibrium by external forcing. Within this setup, we study the lift force experienced by a test bead embedded in a compressible, parity-breaking granular flow. We analyze how odd viscosity and microrotation modify the transverse forces, using both analytical results in the linearized Stokes regime and nonlinear finite element simulations. Our results demonstrate that micropolar fluids provide a physically consistent and symmetry-informed continuum description of chiral granular matter, capable of capturing lift forces that emerge uniquely from odd transport effects.
\end{abstract}

\maketitle
\section{Introduction}



Granular matter, such as familiar substances like sand, grains, and powders, forms a distinct class of materials that behaves differently from traditional solids, liquids, and gases~\cite{Gennes-RevModPhys.71.S374, RevModPhys-Jaeger, andreotti2013granular}. Unlike atomic or molecular substances, which are influenced by thermal motion, granular materials are governed by mechanical interactions, producing significant effects without requiring thermal energy. Composed of macroscopic particles typically larger than \SI{100}{\micro\metre}, granular materials display remarkable properties due to their discrete nature. These include force chains—networks of stress-bearing contacts that dictate how forces propagate through the material~\cite{majmudar2005contact, wang2020connecting, PhysRevLett-Nampoothiri, li2024dynamic}—as well as phenomena like clogging in hoppers~\cite{To-PhysRevLett.86.71} and clustering instabilities in granular gases~\cite{Goldhirsch-PhysRevLett.70.1619}. The interactions between individual particles play a crucial role in these behaviors.

The flow dynamics of granular materials are particularly intriguing, as they exhibit complex rheological behaviors under different conditions~\cite{midi, jop2006constitutive, kamrin_nonlocal_2012, PhysRevLett-Rietz, kou2017granular, shang2024yielding}. In confined spaces or under low shear, they behave like solids, supporting loads and resisting deformation~\cite{Nichol-PhysRevLett.104.078302}. However, when subjected to external forces such as shaking or tilting, they can transition into a fluid-like state, flowing similarly to liquids~\cite{Pouliquen-annurev}. These flows give rise to phenomena such as convection currents, mixing, and segregation. The transition between solid-like and fluid-like states depends not only on external forces but also on particle characteristics such as shape, size, and surface roughness~\cite{Murphy-PhysRevX.9.011014, zhao2023role}. This complexity makes granular flow a rich field of study, with both fundamental scientific significance and practical applications in industries such as pharmaceuticals, agriculture, and construction.

The theory of micropolar fluids, initially proposed in~\cite{eringen_theory_1966}, extends traditional fluid mechanics to incorporate the mechanics of microcontinua (see \cite{lukaszewicz_micropolar_1999} for a review). It specifically considers the angular velocity and rotational inertia of the microstructure at every point within the fluid. One of the most intriguing applications of micropolar fluid mechanics is to characterize granular flows~\cite{lun_kinetic_1991,babic_average_1997,Hayakawa2000,mitarai_collisional_2002,lhuillier_constitutive_2007,saitoh_rheology_2007}. Unlike conventional fluid dynamics, which primarily considers translational motion, the micropolar fluid model integrates the rotational motions of particles, thereby introducing couple stresses and an asymmetric stress tensor into the analysis. This approach is crucial for understanding the interactions within granular flows, where particle rotations play a significant role due to collisions and frictional contacts. As granular materials flow, these microscopic rotations can significantly influence the macroscopic flow properties, leading to phenomena that cannot be described by classical fluid mechanics. Thus, the micropolar fluid model provides a more comprehensive depiction of granular flow, offering deeper insights into their behavior and enabling more accurate predictions of their dynamics in various industrial and natural processes.

More recently, it has been realized that the effects of rotations extend beyond micropolar degrees of freedom. The presence of micropolar degrees of freedom does not necessarily break parity, but if the system responds differently to left- vs. right-handed configurations (i.e., it has a preferred chirality), then parity symmetry is broken. This can happen if the constituents rotate or in the case of granular matter, if the finite-size fluid constituents are mirror asymmetric. Newtonian fluids can exhibit additional transport coefficients known as odd viscosities~\cite{avron_odd_1998,Fruchart2023} when either parity or time-reversal symmetries are broken, as is the case in rotating systems. For example in isotropic planar flows, a single non-dissipative component of the viscosity tensor emerges in systems lacking both time-reversal symmetry and parity. Initially thought to be quite elusive, the importance of odd transport coefficients has grown in research related to planar solids~\cite{scheibner_odd_2020,surowka_odd_2023,ostoja-starzewski_generalizing_2024,fossati_odd_2024,wolfgram_odd_2025}, fluids~\cite{lucas_phenomenology_2014,lingam_action_2014,banerjee_odd_2017,Ganeshan2017,souslov_topological_2019,chattopadhyay_effect_2022,banerjee_hydrodynamic_2022,hosaka_lorentz_2023,gornyi_two-dimensional_2023,machado_monteiro_hamiltonian_2023,poggioli_odd_2023,lier_slip-induced_2024,hosaka_chirotactic_2024,Daddi2025,Franca_Odd_2025}, diffusive systems~\cite{kalz_oscillatory_2024,luigi_muzzeddu_self-diffusion_2025}, liquid crystals~\cite{lingam_hall_2015,pismen_nematodynamics_2024}, and viscoelastic media~\cite{banerjee_active_2021,lier_passive_2022,reichhardt_active_2022,Duclut_Probe_2024,Floyd_Pattern_2024,Matus_Molecular_2024}. In addition, studies of three-dimensional fluids were also performed~\cite{markovich_odd_2021,khain_stokes_2022,reynolds_hele-shaw_2022,lier_odd_2024,everts_dissipative_2024,khain_trading_2024,Matus:2024kyg}. In this case, the number of coefficients increases due to the breaking of isotropy by parity-odd shapes. Odd-transport-related phenomena have been proposed to exist in certain active or quantum materials, leading to experimental realizations in colloidal~\cite{soni_odd_2019}, electronic systems~\cite{berdyugin_measuring_2019}, living matter~\cite{tan_odd_2022}, and wood~\cite{ozyhar_viscoelastic_2013}.

Since granular materials naturally incorporate the importance of rotations, including odd viscosities in chiral granular matter is essential to accurately account for the symmetries of these systems. Meanwhile, chiral transport in planar granular flows has not yet been explored. In order to remedy this, in this work, we begin to integrate odd viscosity into the flows of micropolar fluids. Specifically, we investigate the phenomenology of a bead embedded within compressible micropolar fluids, marking an initial step towards understanding the impacts of odd viscosity in such systems. Our primary focus is on the phenomenology of lift on a test bead. 

Our primary motivation is driven by the potential realization of chiral micropolar flows, which are based on vibrated discs~\cite{deseigne_collective_2010,chen_high-energy_2022,chen_anomalous_2024}. As a result, in addition to applying analytical methods developed for linearized Stokes fluids in infinite domains, we also focus on numerical finite element methods that allow us to study fully nonlinear equations in finite-sized channels. By corroborating the analytical results within their domain of validity, we expect that the numerical approach will be valid for describing chiral flows under realistic experimental conditions. Our main focus is the lift force experienced by a test bead immersed in a flow of chiral granular matter confined within a finite-size domain, as in typical experimental setups. In such a system, chiral particles are restricted to a bounded region, where we embed a test object and induce its motion relative to the surrounding medium. By driving the object through the chiral granular flow, we aim to observe and analyze the resulting lift force, which arises due to the interplay between the broken symmetries of the medium and the relative motion of the bead.

In this work, we develop a framework for \emph{driven micropolar fluids} to describe vibrated granular layers. The systems of interest are maintained in non-equilibrium steady states through continuous external energy injection, rather than relaxing to equilibrium once the driving is removed. Our motivation is to model macroscopically vibrated discs, where sustained forcing compensates for dissipative particle interactions. Within this symmetry-based formulation, we include only those terms permitted by the underlying symmetries and do not introduce transport coefficients forbidden by the second law of thermodynamics. The external driving and dissipation are thus encoded effectively in the values of the transport coefficients, which need not obey equilibrium constraints.

\section{Hydrodynamic Framework for Parity-Breaking Granular Fluids}
Hydrodynamics offers a low-energy description of the behavior of interacting many-body systems. It focuses on a specific set of physical quantities, such as particle number and momentum, which are conserved and thus play a crucial role at low energies. In the most economical formulation, the behavior of a fluid can be described using a velocity field \(v_i\) and two thermodynamic variables, which in our case will be pressure \(P\) and density \(\rho\). An extension of this minimal framework to account for rotations and chirality demands the introduction of a new field \(\xi\), whose role is to capture the internal rotations of the fluid constituents.

Granular matter consists of large collections of solid particles and exhibits behavior fundamentally distinct from that of conventional solids or liquids. Most existing models for granular materials are phenomenological, primarily aimed at addressing engineering applications, while many fundamental mechanical questions remain unresolved. Granular matter is inherently multiscale, comprising the microscale of individual particles, the mesoscale of force chains, and the macroscale of the material bulk. Interactions across these scales are crucial to understanding its behavior.

One may ask whether it is possible to apply hydrodynamic equations to granular materials by defining local variables such as density, velocity, and an effective temperature. However, these systems dissipate energy rapidly and often fail to reach equilibrium, a key assumption in conventional hydrodynamics. Phenomena such as inelastic collapse, clustering, shear bands, and history-dependent compaction challenge the notion of a universal, local, and time-independent hydrodynamic theory. Simulations and experiments show that while some aspects of granular flow can be captured by modified hydrodynamic-like equations, a complete description typically requires incorporating nonlocal effects, stochastic dynamics, and microscopic considerations beyond standard hydrodynamic models. At the same time, the hydrodynamics of active matter offers promising avenues for extending conventional hydrodynamic equations to non-equilibrium systems. By incorporating internal driving forces and accounting for persistent energy input at the microscopic level, active matter hydrodynamics demonstrates that it is possible to develop continuum theories that remain predictive even far from equilibrium. This progress suggests that, with appropriate modifications, hydrodynamic frameworks could be generalized to describe complex systems like granular flows, despite their inherent dissipation and lack of equilibrium.

In vibrated or sheared granular systems, steady flows emerge from a dynamic balance between energy injection and dissipation, see e.g.~\cite{lhuillier_constitutive_2007,kamrin_nonlocal_2012}. Unlike molecular fluids, granular materials lose energy through inelastic collisions, which makes continuous external forcing essential to sustain motion. When the energy supplied to the system compensates for the dissipative losses, the system can reach a steady state characterized by stable macroscopic fields. These steady states are not only accessible through theoretical models but are also well documented in experimental setups involving vertically vibrated beds and horizontally driven flows~\cite{goldhirsch_r_2003,aranson_patterns_2006}.

Micropolar fluid theory is a natural extension of classical hydrodynamics and is particularly well-suited for capturing the multiscale characteristics of dense granular matter. Unlike conventional fluids, micropolar fluids incorporate additional, gapped degrees of freedom, including local particle rotations and couple stresses, which are essential for describing systems where particle-scale effects and internal structure significantly influence macroscopic behavior. This theoretical framework allows for the representation of mesoscale features such as rotational inertia and anisotropic stress transmission that arise from collective particle interactions. Although it does not explicitly resolve individual force chains, micropolar theory accounts for their averaged effects and captures the role of internal microstructure in shaping the overall mechanical response. By including particle rotations and the torques they generate, it offers a more accurate and physically consistent continuum description of dense granular flows, especially in regimes where classical hydrodynamics falls short due to its inability to represent nonlocal interactions and internal degrees of freedom.

We incorporate external force \(f_i\) and torque density \(g\) acting on the granular layer. Together, these modifications enable the model to more accurately describe the irreversible and non-equilibrium behavior of dense granular flows in contact with a larger system. By incorporating this symmetry breaking, the extended framework more accurately represents the irreversible and non-equilibrium behavior of dense granular flows and provides a more realistic continuum description of their dynamics.
Taking the above phenomenology into account we propose the governing equations to be the conservation of mass
\begin{equation}
    \partial_t \rho + \partial_k (\rho v_k) = 0,
    \label{eq:cons_mass}
\end{equation}
conservation of momentum
\begin{equation}
    \rho ( \partial_t + v_k \partial_k ) v_j = \partial_i T_{ij} + f_j,
    \label{eq:cons_momentum}
\end{equation}
and conservation of angular momentum
\begin{equation}
    \rho I ( \partial_t + v_k \partial_k ) \xi = \partial_i C_i + \epsilon_{ij} T_{ij} + g,
    \label{eq:cons_ang_momentum}
\end{equation}
where $I$ is a microinertia coefficient -- and an equation of state (EoS) $P(\rho)$ which for a weakly compressible fluid takes form of 
\begin{equation}
    P = P_0 + \chi \frac{ \rho - \rho_0 }{ \rho_0 },
    \label{eq:eo_state}
\end{equation}
where $P_0$ and $\rho_0$ describe reference state of the fluid and coefficient $\chi^{-1}$ is the compressibility. $T_{ij}$ add $C_i$ represent fluid's stress tensor and couple stress tensor, while $\epsilon_{ij}$ denotes the standard two-dimensional Levi-Civita antisymmetric tensor.

We note that in this work, we do not aim to construct a detailed constitutive theory derived from kinetic considerations~\cite{brilliantov_kinetic_2004}, but instead adopt a macroscopic continuum framework based on symmetry principles that naturally lead to micropolar fluid theory. Micropolar models for granular media have been justified using also coarse-graining procedures~\cite{artoni_coarse_2019}, based on developments in~\cite{artoni_average_2015}. Within our framework, we incorporate odd viscosity terms, which are consistent with broken parity and are essential for capturing chiral effects such as the lift force investigated here. To isolate and understand the minimal conditions under which such lift can arise, we consider constant transport coefficients. While in rapid granular flows these coefficients generally depend on granular temperature and energy dissipation, such dependencies represent transient corrections beyond the steady-state regime considered here. In fact energy balance is not the only correction that could impact transient phenomena in granular matter. Various rheological responses should also play an important role. This behavior is analogous to viscoelastic extensions of Newtonian fluids~\cite{Duclut_Probe_2024}, where transient corrections emerge from internal relaxation mechanisms. In this light, granular matter may be viewed as an active generalization of micropolar fluids, incorporating both rheological complexity and symmetries.

In the present discussion, we focus on media that violate parity. They exhibit an intriguing characteristic where the typical symmetry associated with mirror reflections is absent at a microscopic level. In fluids this asymmetry can be attributed to external forces, like magnetic fields, or can arise from inherent activities within the fluid, such as the exertion of microscopic torques. In granular matter, due to the finite size of the constituents, parity symmetry can be broken by the constituents themselves. An example is given by a a rattleback, also known as a celt or wobblestone~\cite{Garcia1988,Zhuravlev2008}. A rattleback is a semi-ellipsoidal top which spins on a flat surface, but exhibits the unusual behavior of spinning preferentially in one direction. This directional preference and the resulting reversal in spin are due to the breaking of parity symmetry in its physical design and mass distribution. The rattleback's asymmetry isn't just in its shape—it also involves how mass is distributed within the object. Typically, the center of mass is not aligned with the geometric center, and the principal axes of inertia are not aligned symmetrically with the base. In modern experiments chiral objects can also be constructed in more controlled ways; chirality is introduced by attaching asymmetric legs beneath rotating disks. The goal of this work is to investigate a macroscopic, hydrodynamic description of a granular fluid made of such circular, parity-breaking constituents on a plane.
In order to arrive at a closed system of equations we need the constitutive relations between currents and fields. Since parity is broken the most general form reads
\begin{equation}
     T_{ij} = 2\eta_{s} \partial_{\{j} v_{i\}} + 2\eta_o \partial_{\langle j}  v_{i\rangle}+\delta_{ij}(\eta_b\partial_kv_k-P)+\mu_r((\partial_i v_j-\partial_jv_i)+2\epsilon_{ji}\xi), 
     \label{eq:stress_tensor}
\end{equation}
\begin{equation}
     C_i=c_1\partial_i\xi,
     \label{eq:couple_stress_tensor}
\end{equation}
where $\eta_s$, $\eta_o$, $\eta_b$ and $\mu_r$ denote respectively shear, odd, bulk and dynamic microrotation viscosities and $c_1$ is a coefficient of angular viscosity. Additionally we use $A_{\langle ij\rangle}=(A_{ij}+A_{ji})/2-A_{kk}\delta_{ij}/2$ and $A_{\{ ij\}}=(\epsilon_{ik}A_{jk}+\epsilon_{ik}A_{kj}+\epsilon_{jk}A_{ik}+\epsilon_{jk}A_{ki})/4$, with $\epsilon_{ij}$ being the Levi-Civita tensor.

\section{Forces on a Body in a Micropolar Brinkman-Type Fluid}

The Stokes problem concerning a sphere moving in a medium, often referred to in fluid dynamics as ``Stokes flow past a sphere,'' is a classic problem that involves analyzing the behavior of a fluid flowing around a sphere in relative motion. This scenario is particularly relevant in the low--Reynolds-number regime, where viscous forces dominate over inertia. To address the analogous situation of a body moving through a granular medium, it is necessary to determine the motion of the particle as it responds to specific forces and torques within an ambient flow. We use the Stokes limit not as a literal description of granular rheology, but as a toy model, in which the lift effect can be isolated analytically. This approach is common in active-matter and complex-fluid contexts, where it provides transparent insight into how symmetry breaking, here, due to chirality and odd viscosity, modifies hydrodynamic forces. The principal result, namely the existence of a lift force arising from the combined effects of odd viscosity and bead chirality, remains qualitatively robust even when full nonlinear dynamics are retained, as confirmed by our numerical simulations.  

Hydrodynamic evolution equations are intrinsically nonlinear due to the presence of terms involving products of velocity components and their derivatives. To facilitate analytic progress and obtain closed-form expressions for response functions, we employ a linearization technique. This method involves expanding the equations to first order in the perturbation variables \(v_i\), \(\xi\), and \(\delta\rho = \rho - \rho_0\), around a state characterized by negligible velocities and a homogeneous reference state. In this \emph{linearized setting}, we introduce phenomenological relaxation terms into the micropolar hydrodynamic equations.

It is convenient to view this regularized linearized framework as a \emph{toy Brinkman model} for a chiral micropolar fluid. The Brinkman equation \cite{brinkman_calculation_1949}, which extends Stokes flow by incorporating an effective drag term proportional to velocity, is widely used as a minimal model for flow in porous or frictional environments. Analogously, the present formulation captures the essential balance between viscous stresses, odd and rotational couplings, and environmental dissipation, while retaining analytical tractability. The parameters \(\kappa\) and \(\tau\) thus play the role of phenomenological screening lengths that control the exponential decay of perturbations, rather than representing intrinsic microscopic processes.  

In this spirit, the resulting linearized equations constitute a \emph{toy Brinkman-type model} for a chiral micropolar fluid and take the form
\begin{equation}
    \partial_t\delta\rho+\rho_0\partial_k v_k = -\frac{1}{\kappa}\delta\rho,
    \label{eq:cons_mass_lin}
\end{equation}
\begin{equation}
    \rho_0\partial_t v_j=-\partial_j P+(\eta_s+\mu_r)\Delta v_j+(\eta_b-\mu_r)\partial_j\partial_i v_i+\eta_o\epsilon_{j i}\Delta v_i-\frac{\rho_0 v_j}{\tau}+2\mu_r\epsilon_{j i}\partial_i\xi+f_j,
    \label{eq:cons_momentum_lin}
\end{equation}
\begin{equation}
    2\mu_r\epsilon_{i j}\partial_i v_j = \rho_0 I\partial_t\xi-c_1\Delta\xi+4\mu_r\xi+\frac{\rho_0 I}{\alpha}\xi-g.
    \label{eq:cons_ang_momentum_lin}
\end{equation} 

If a nontrivial lift force arises already within this simplified Brinkman-type model, where inertial and nonlinear convective terms are suppressed, its origin must lie in the symmetry structure of the constitutive relations themselves rather than in any specific nonlinear mechanism. Consequently, the presence of an odd-viscosity-induced lift force in this linear regime guarantees that the same effect will persist qualitatively in the fully nonlinear flow, where it may only be quantitatively renormalized by higher-order couplings and boundary effects.  

To analyze this mechanism analytically, it is convenient to work in Fourier space using the Fourier transform $g(\omega,k_i)=\int dt \, d^2 x_i  \, g(t, x_i) e^{i\omega t- ik_j x_j}$, which is natural for a spatially homogeneous and translationally invariant bulk fluid. This representation diagonalizes the linearized operators and allows for the explicit evaluation of the response and lift coefficients. This approach follows standard practice in linearized hydrodynamics \cite{landau_fluid_1987,hinch_perturbation_1991,pozrikidis_boundary_1992}, where the Fourier representation provides a natural framework for deriving response functions and identifying symmetry-protected transport effects.

We rewrite Eq.~(\ref{eq:cons_mass_lin}) as
\begin{equation}
    \frac{\delta\rho}{\rho_0}=\frac{-i\kappa k_j}{1-i\omega\kappa}v_j,
    \label{eq:fourier_mass}
\end{equation}
which allows for pressure to be solved in terms of velocity. In the context of low Reynolds number flow, the \emph{resistance matrix} and the \emph{mobility matrix} are crucial concepts to describe the relationship between forces and motions of solid bodies in a viscous fluid~\cite{kim_microhydrodynamics:_2013}. Using Eqs.~(\ref{eq:cons_momentum_lin}) and (\ref{eq:cons_ang_momentum_lin}) we can express the resistance matrix as follows
\begin{equation}
    \begin{pmatrix}
        f_i \\ g
    \end{pmatrix} =
    \begin{pmatrix}
        \mathcal{A}_{ij} & \mathcal{B}_{i}\\
        \mathcal{B}_{j}^* & \mathcal{D}
    \end{pmatrix}
    \begin{pmatrix}
        v_j \\ \xi
    \end{pmatrix}=\mathbf{R}\begin{pmatrix}
        v_j \\ \xi
    \end{pmatrix},
    \label{eq:resistance_matrix}
\end{equation}
where
$$\mathcal{A}_{ij} = \left(\frac{\rho_0}{\tau}-i\omega\rho_0+(\eta_s+\mu_r)k^2\right)\delta_{ij}
  +\left((\eta_b-\mu_r)k^2+\frac{\chi\kappa k^2}{1-i\omega\kappa}\right)\hat{k}_i\hat{k}_j+\eta_o k^2\epsilon_{ij},$$
$$\mathcal{B}_{i} = -2\mu_ri\epsilon_{ij}k_j,$$
$$\mathcal{D} = -i\rho_0 I\omega+c_1k^2+4\mu_r+\frac{\rho_0 I}{\alpha} .$$
The inverse of $\mathbf{R}$, gives the mobility matrix
\begin{equation}
    \mathbf{M} =(\mathcal{D}\mathcal{A}_{ij}-\mathcal{B}_i\mathcal{B}_j^*)^{-1}
    \begin{pmatrix}
        \mathcal{D} & -\mathcal{B}_i\\
        -\mathcal{B}_j^* & \mathcal{A}_{ij}
    \end{pmatrix} = \mathbf{R}^{-1}.
    \label{eq:mobility_matrix}
\end{equation}
Instead of computing the velocity field in the surrounding fluid it is convenient to transform the Stokes equations into an integral form that is applied directly over the surface of the object. The boundary conditions on the surface of an object can be viewed as applying forces to the surrounding fluid, altering the fluid's flow patterns around the object. By representing the object with a collection of force singularities this method effectively replicates the boundary conditions. This allows us to directly address the mobility problem by modeling how these forces influence the fluid dynamics. Moreover, in the Fourier space a technical simplification occurs, which facilitates the computation of required integrals. This is known as the shell localization method. We decompose external force and torque densities as $f_i=L(k)\mathcal{F}_i(\omega)$ and $g=L(k)\gamma(\omega)$. Since we consider a cylindrical bead of radius $a$ as it was done by \cite{lift_odd_compr} we will set $L(k)=J_0(ak)$, where $J_n$ is the $n$-th Bessel function of first kind. To obtain an expression for the velocity and rotation of the disk we calculate
\begin{equation}
    \begin{pmatrix}
        v_j \\ \xi
    \end{pmatrix} (\omega,|x|=0)= \frac{1}{(2\pi)^2}\int_0^{2\pi}d\theta\int_0^\infty dk
    J_0(ak)
    \mathbf{M} 
    \begin{pmatrix}
       \mathcal{F}_i(\omega) \\ \gamma(\omega)
    \end{pmatrix}.
    \label{eq:vel_at_origin}
\end{equation}
In our study focused on quantifying the lift and drag forces acting on the bead, we selectively address one of the derived equations critical to our analysis:
\begin{equation}
    v_i(\omega,|x|=0)=\mathbb{M}_{ij}(\omega)\tilde{\mathcal{F}}_j(\omega),
    \label{eq:velocity_disk}
\end{equation}
where $\tilde{\mathcal{F}}_i(\omega)=\mathcal{F}_i(\omega)-\mathcal{B}_i\mathcal{D}^{-1}\gamma(\omega)$ and
\begin{equation}
  \mathbb{M}_{ij}(\omega)=\frac{1}{(2\pi)^2}\int_0^{2\pi}d\theta\int_0^\infty dk \,kJ_0(ak)\left(\mathcal{A}_{ij}-\mathcal{B}_i\mathcal{B}_j^*\mathcal{D}^{-1}\right)^{-1}  
  \label{eq:response_matrix}
\end{equation}
is the "response matrix" encoding the velocity of the cylindrical bead immersed in the fluid as a function of applied frequency-dependent force $\tilde{\mathcal{F}}_j(\omega)$. Based on symmetry considerations we decompose the response matrix as follows
\begin{equation}
    \mathbb{M}_{ij}=\frac{1}{\eta_s}(M_d\delta_{ij}-M_l\epsilon_{ij}),
    \label{eq:response_drag_lift}
\end{equation}
with $M_d$ and $M_l$ being the dimensionless response coefficients for drag and lift force respectively.

For subsequent calculations, we introduce the following set of dimensionless quantities
\begin{align*}
z_i&=ak_i,   &  
\bar{\omega}&=\omega\frac{\rho_0 a^2}{\eta_s},         &  
\bar{\eta}_o&=\frac{\eta_o}{\eta_s}, & 
\bar{\eta}_b&=\frac{\eta_b}{\eta_s}, &
\bar{\tau}&=\tau\frac{\eta_s}{\rho_0 a^2},       &  \bar{\chi}&=\chi\frac{\rho_0 a^2}{\eta_s^2},   &  \\\bar{\kappa}&=\kappa\frac{\eta_s}{\rho_0 a^2}, &  \bar{\mu}_r&=\frac{\mu_r}{\eta_s}, & \bar{I}&=\frac{I}{a^2}, &\bar{c}_1&=\frac{c_1}{a^2\eta_s}, & \bar{\alpha}&=\alpha\frac{\eta_s}{\rho_0 a^2}.
\end{align*}

Coefficients $M_d$ and $M_l$ can be written explicitly: in the integral form
\begin{equation}
    M_d=\frac{1}{4\pi}\int_0^\infty dzJ_0(z)z\frac{2A+B}{A^2+AB+C^2},
    \label{eq:drag_coeff}
\end{equation}
\begin{equation}
    M_l=\frac{1}{2\pi}\int_0^\infty dzJ_0(z)z\frac{C}{A^2+AB+C^2},
    \label{eq:lift_coeff}
\end{equation}
where we have defined
\begin{align*}
    A(z)&=\bar{\tau}^{-1}-i\bar{\omega}+z^2\left(1+\bar{\mu}_r-\frac{4\bar{\mu}_r^2}{\bar{c}_1z^2+4\bar{\mu}_r+\bar{I}(\bar{\alpha}^{-1}-i\bar{\omega})}\right), \\
    B(z)&=z^2\left(\bar{\eta}_b-\bar{\mu}_r+\frac{\bar{\chi}\bar{\kappa}}{1-i\bar{\omega}\bar{\kappa}}+\frac{4\bar{\mu}_r^2}{\bar{c}_1z^2+4\bar{\mu}_r+\bar{I}(\bar{\alpha}^{-1}-i\bar{\omega})}\right),\\
C(z)&=\bar{\eta}_oz^2.
\end{align*}

The momentum integrals required for the evaluation of $M_d$ and $M_l$ can be computed analytically by employing the residue theorem~\cite{Lin2013}. Before we embark on numerical techniques, we will present two analytical examples of solutions. The first parallels the classical steady-state problem, where the flow does not change over time. The second concerns an oscillatory flow, in which either the bead or the flow conditions for the medium vary sinusoidally with time. A key insight emerges from the analytical structure of the lift force: as evident from Eq.~(\ref{eq:lift_coeff}) (see also Fig.~\ref{fig:linearization}), in the absence of odd viscosity, the lift component $M_l$ vanishes—recovering the familiar case where only drag is present. This highlights the novel role of odd viscosity, which, unlike conventional viscosity, is non-dissipative. Its effect is analogous to that of a magnetic field acting on a charged particle. Odd viscosity enters the stress tensor through antisymmetric derivative couplings, producing a transverse force perpendicular to the local velocity gradients. This mechanism is formally analogous to the Lorentz force, where a magnetic field couples antisymmetrically to the velocity, generating a transverse current. In both cases, the effect preserves energy (since the stress tensor remains traceless in the odd sector, just as the Lorentz force does no work), but it modifies the momentum balance by deflecting flows.

\subsection{Steady state}
\begin{figure}
    \centering
    \includegraphics[width=0.9\linewidth]{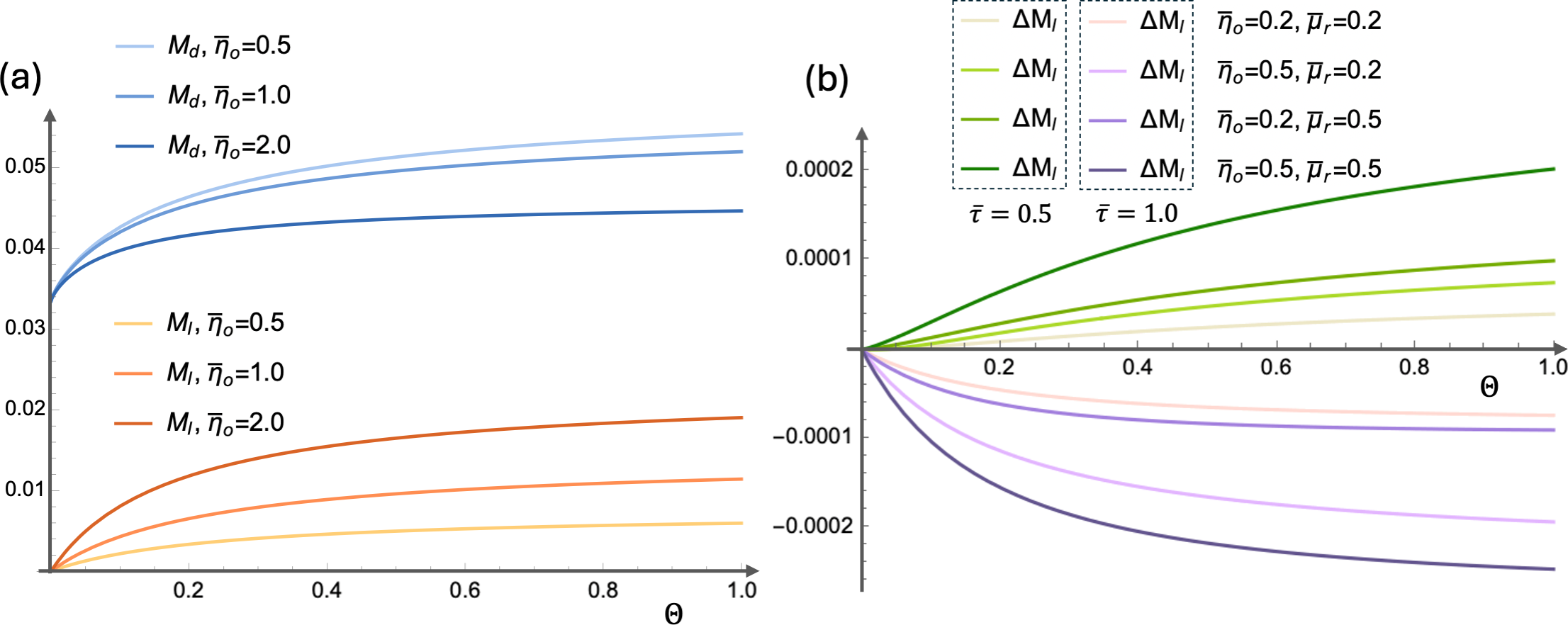}
    \caption{Steady state analytical solutions for (a) drag $M_d$ and lift $M_l$ coefficients, (b) correction to the lift coefficient $\Delta M_l$ due to microrotation. Unless otherwise specified, the parameters take the following values: $\bar{\tau}=1$, $\bar{\eta}_b=1$, $\bar{\mu}_r=0.4$, $\bar{I}=0.1$, $\bar{c}_1=2$, $\bar{\alpha}=0.5$.}
    \label{fig:linearization}
\end{figure}
To understand the forces on a bead involving microscale rotational effects, we first consider how microrotation influences the steady-state behavior of the system, characterized by conditions where $\bar{\omega} \rightarrow 0$ (indicating non-oscillatory behavior), while $\bar{\tau}^{-1}$ and $\bar{\alpha}^{-1}$ remain significant, affecting the fluid’s response. Example steady-state solutions of Eqs.~\ref{eq:drag_coeff} and~\ref{eq:lift_coeff} are presented in Fig.~\ref{fig:linearization} showing (a) $M_d$ and $M_l$ as a function of $\Theta=(\bar{\chi}\bar{\kappa})^{-1}$, and (b) correction to the lift force due to microrotation, defined as $\Delta M_l = M_l-M_l(\mu_r=0)$.

In our next analysis, we simplify the computational process by calculating the coefficients $M_d$ (drag) and $M_l$ (lift) as expansions in terms of the odd viscosity $\bar{\eta}_o$, retaining only the first non-vanishing term. This approach allows us to efficiently capture the primary effects of microrotation on the fluid dynamics under steady-state conditions. We obtain the following drag and lift coeffients:
\begin{equation}
    M_d=\frac{2K_0[(\Xi\bar{\tau})^{-1/2}]/\Xi+\frac{1+\Phi}{1+\bar{\mu}_r}K_0(\sqrt{\Pi^{+}})+\frac{1-\Phi}{1+\bar{\mu}_r}K_0(\sqrt{\Pi^{-}})}{8\pi}+\mathcal{O}(\bar{\eta}_o^2) ,
    \label{eq:md_steady}
\end{equation}
 \begin{equation}
    M_l= \frac{\bar{\eta}_o}{4\pi}\left(\frac{-2K_0[(\Xi\bar{\tau})^{-1/2}]/\Xi}{\Omega+\Xi-(1+\bar{\mu}_r)} +\frac{\frac{1+\Phi}{1+\bar{\mu}_r}K_0(\sqrt{\Pi^{+}})}{\Xi-(\bar{\tau}\Pi^+)^{-1}}+\frac{\frac{1-\Phi}{1+\bar{\mu}_r}K_0(\sqrt{\Pi^{-}})}{\Xi-(\bar{\tau}\Pi^-)^{-1}}\right)+\mathcal{O}(\bar{\eta}_o^2),
    \label{eq:ml_steady}
\end{equation}
where
\begin{align*}
    \Phi&=\frac{\frac{1}{2}b-\frac{4\bar{\mu}_r+\frac{\bar{I}}{\bar{\alpha}}}{\bar{c}_1 }}{\sqrt{\frac{1}{4}b^2-\frac{4\bar{\mu}_r+\frac{\bar{I}}{\bar{\alpha}}}{\bar{c}_1 \bar{\tau}(1+\bar{\mu}_r)}}}, &  \Pi^\pm &= \frac{1}{2}b\pm\sqrt{\frac{1}{4}b^2-\frac{4\bar{\mu}_r+\frac{\bar{I}}{\bar{\alpha}}}{\bar{c}_1 \bar{\tau}(1+\bar{\mu}_r)}}, \\
    b&=\frac{4\bar{\mu}_r+\frac{\bar{I}}{\bar{\alpha}}(1+\bar{\mu}_r)}{\bar{c}_1(1+\bar{\mu}_r)}+\frac{\bar{\tau}^{-1}}{1+\bar{\mu}_r}, & \Omega&=\frac{4\bar{\mu}_r^2}{4\bar{\mu}_r+\frac{\bar{I}}{\bar{\alpha}}-\bar{c}_1(\bar{\tau}\Xi)^{-1}},
\end{align*}
are functions of the parameters that receive functional dependence on the microrotational viscosity, and
$$ \Xi=1+\bar{\eta}_b+\bar{\chi}\bar{\kappa} $$
is a function that captures the compressibility of the medium.

Equations (\ref{eq:md_steady}) and (\ref{eq:ml_steady}) delineate a rather complex relation between $M_l$, $M_d$ and $\bar{\mu}_r$. Notably, as $\bar{\mu}_r$ approaches zero, both $M_d$ and $M_l$ asymptotically approach their respective forms in the absence of microrotation~\cite{lift_odd_compr}.  This behavior is expected and serves as an important cross-check with previous results, confirming that as microrotational viscosity diminishes, the velocity field and microrotation effectively decouple, reverting to a classical non-microrotational dynamic.

For small values of $\bar{\mu}_r$ equations for the response coefficients take form
\begin{equation}
    M_d=M_d^0+\frac{K_1[\bar{\tau}^{-1/2}]\bar{\tau}^{-1/2}-2K_0[\bar{\tau}^{-1/2}]}{8\pi}\bar{\mu}_r+\mathcal{O}(\bar{\eta}_o^2,\bar{\mu}_r^2), 
    \label{eq:md_steady_fo}
\end{equation}
\begin{equation}
    M_l= M_l^0+\frac{\frac{K_1[\bar{\tau}^{-\frac{1}{2}}]\bar{\tau}^{-\frac{1}{2}}}{\Xi-1}-\frac{2K_0[(\Xi\bar{\tau})^{-\frac{1}{2}}]/\Xi}{(\Xi-1)^2} -\frac{2K_0[\bar{\tau}^{-\frac{1}{2}}](\Xi-2)}{(\Xi-1)^2}}{4\pi}\bar{\eta}_o\bar{\mu}_r+\mathcal{O}(\bar{\eta}_o^2,\bar{\mu}_r^2).
    \label{eq:ml_steady_fo}
\end{equation}
We can see that the first order correction to the drag force only depends on the value of momentum relaxation whereas lift force heavily depends on compressibility as well. It can also be seen that as either $\bar{\kappa}$ or $\bar{\chi}$ approach infinity (case of an incompressible fluid or a fluid without mass relaxation) -- lift force disappears and $M_d=K_0(\bar{\tau}^{-1})/(4\pi)$. This is also consistent with previous studies~\cite{lift_odd_compr}. Later on, we will numerically (FEM) solve nonlinear equations for compressible fluids, where mass relaxation is effectively encapsulated by the nonlinear terms.

\subsection{Frequency-dependent lift force}

Now we shall consider a limit in which the relaxation process is absent i.e. $\bar{\tau}^{-1}\rightarrow 0$, $\bar{\kappa}^{-1}\rightarrow 0$ and $\bar{\alpha}^{-1}\rightarrow 0$. By expanding in terms of the inverse of compressibility $\bar{\chi}^{-1}$ we can obtain simple analytical solutions given by: 
\begin{equation}
    M_d=\frac{1}{8\pi}\left(\frac{1+\Psi}{1+\bar{\mu}_r}K_0(\sqrt{\Sigma^+})+\frac{1-\Psi}{1+\bar{\mu}_r}K_0(\sqrt{\Sigma^-}) \right)+\mathcal{O}(\bar{\chi}^{-1}),
    \label{eq:md_freq}
\end{equation}
\begin{equation}
    M_l=\frac{-i\bar{\omega}\bar{\eta}_o}{4\pi\bar{\chi}}\left(\frac{1+\Psi}{1+\bar{\mu}_r}K_0(\sqrt{\Sigma^+})+\frac{1-\Psi}{1+\bar{\mu}_r}K_0(\sqrt{\Sigma^-}) \right)+\mathcal{O}(\bar{\chi}^{-2}),
    \label{eq:ml_freq}
\end{equation}
where
$$
    \Psi=\frac{\frac{1}{2}s-\frac{4\bar{\mu}_r-i\bar{\omega}\bar{I}}{\bar{c}_1}}{\sqrt{\frac{1}{4}s^2+i\bar{\omega}\frac{4\bar{\mu}_r-i\bar{\omega}\bar{I}}{\bar{c}_1(1+\bar{\mu}_r)}}}, $$$$
    \Sigma^\pm = \frac{1}{2}s\pm\sqrt{\frac{1}{4}s^2+i\bar{\omega}\frac{4\bar{\mu}_r-i\bar{\omega}\bar{I}}{\bar{c}_1(1+\bar{\mu}_r)}}, $$$$
      s=\frac{4\bar{\mu}_r-i\bar{\omega}\bar{I}(1+\bar{\mu}_r)}{\bar{c}_1(1+\bar{\mu}_r)}-\frac{i\bar{\omega}}{1+\bar{\mu}_r} 
$$
are again functions that receive corrections representing the impact of microrotation on this system. For small values of $\bar{\mu}_r$:
\begin{equation}
    M_d=M_d^0+\frac{\bar{\mu}_r}{8\pi}\left(K_1[\sqrt{\bar{\omega}/i}]\sqrt{\bar{\omega}/i}-2K_0[\sqrt{\bar{\omega}/i}]\right)+\mathcal{O}(\bar{\chi}^{-1},\bar{\mu}_r^2),
    \label{eq:md_freq_fo}
\end{equation}
\begin{equation}
    M_l=M_l^0-\frac{i\bar{\omega}\bar{\eta}_o\bar{\mu}_r}{4\pi\bar{\chi}}\left(K_1[\sqrt{\bar{\omega}/i}]\sqrt{\bar{\omega}/i}-2K_0[\sqrt{\bar{\omega}/i}]\right)+\mathcal{O}(\bar{\chi}^{-2},\bar{\mu}_r^2).
    \label{eq:ml_freq_fo}
\end{equation}

\section{Finite domain calculations}\label{sec:num}

Having identified the lift effect within the \emph{toy Brinkman-type linearized model, we now investigate its manifestation in the full nonlinear compressible Navier–Stokes system Eqs.~(\ref{eq:cons_mass}-\ref{eq:couple_stress_tensor})}. To get numerical solutions to the system we will use finite element method (FEM) with a proper variational formulation that gives discretization of the continuous differential system on the grid that is adapted to the geometry of our problem. In the presented calculations we have used a simple but efficient \textit{splitting method} also known as \textit{Chorin method}~\cite{Chorin1968} or \textit{incremental pressure correction scheme} (IPCS)~\cite{Goda1979}. IPCS is typically used for finding stationary solutions for incompressible fluids, but here, by using some improvements, we were able to adapt it to odd compressible fluid coupled with a microrotation field. Details about the variational formulation of the used FEM as well as detailed formulation of the numerical iterative procedure can be found in the Appendix.

\begin{figure}
    \centering
    \includegraphics[width=0.9\linewidth]{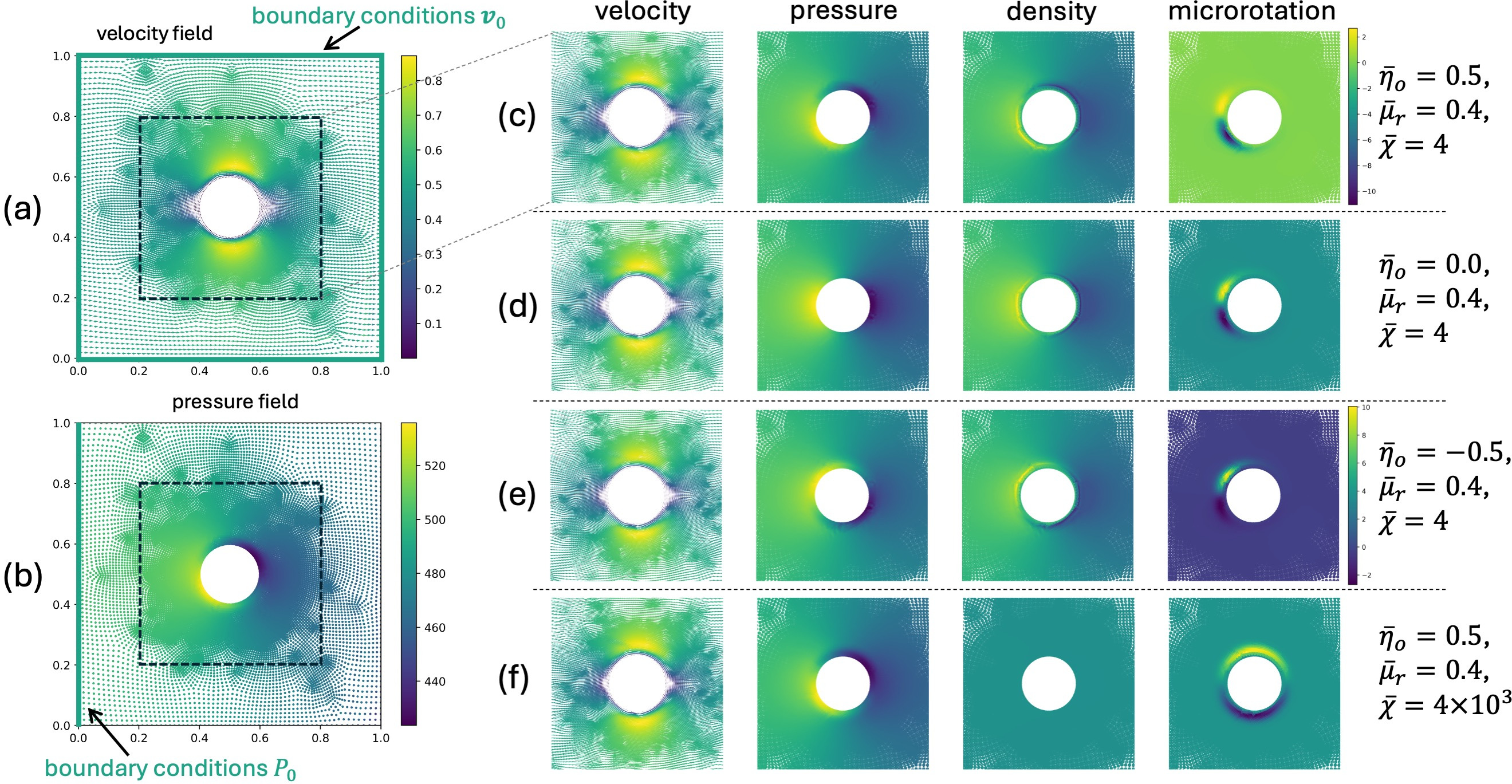}
    \caption{Finite element method results for a compressible odd fluid coupled to a micropolar field flowing through a bead disk in a finite domain are presented. Panels (a,b) show the computational domains (grid is also shown) and boundary conditions: (a) the velocity field with its boundary condition (b.c.) $\bm{v}_0$ (marked by green square), and (b) the pressure field with pressure b.c. $P_0$ applied along the left wall (green segment). Example solution (velocity, pressure, density, and microrotation) fields are displayed for fluids with (c) positive oddity, (d) zero oddity, and (e) negative oddity. For comparison, panel (f) uses the same parameters as (c) but for an incompressible fluid.}
    \label{fig:fem_example}
\end{figure}
Before we start with the FEM results, let us also comment on two important differences between the problem definition in analytical and numerical domains.
Due to the fact that in numerics we have to deal with a finite area, i.e. a computational box that encloses the cylindrical bead, in contrast to analytical domain where we solved the equations for infinite surrounding of 2D bead.
To define a problem in a finite area we have to setup the proper boundary conditions. In Fig.~\ref{fig:fem_example}(a) there is presented computational box for the velocity field $v_i$ with the applied constant velocity $\bm{v}^\mathrm{b}=[v_0,0]$ at the edges (marked by a green box). Moreover, to couple the bead with the fluid we apply no-slip boundary conditions at the bead edge $\Gamma^\mathrm{p}$, visible as velocity field vanishing close to the central disk (the bead), $\bm{v}_\parallel=0$ at $\Gamma_\mathrm{p}$, in Fig.~\ref{fig:fem_example}(a). Also, to setup the pressure offset level $P_0$ in Eq.~(\ref{eq:eo_state}), at the left wall of the computational box (marked by a green section) for the pressure field shown in Fig.~\ref{fig:fem_example}(b), there is set the $P_0$ boundary condition.

The finite velocity $\bm{v}^\mathrm{b}$ at the top and bottom boundary of the computational box makes that our numerical formulation resembles rather \textit{Poiseuille flow} through a rectangular pipe with an additional cylindrical obstacle rather than observation of a force density $f_i$ introduced by a cylindrical tracer dragged with some velocity through the fluid (in a steady state the velocity saturates to $\tau f_i$). To reconcile these two formulations, we change coordinates to the resting bead and observe forces acting on its edge by the fluid. To this end, we assume that in the numerical formulation, neither external forces, i.e. $f_i=0$, nor saturation term, i.e. $\tau\rightarrow\infty$, are present. However, the equivalent of saturation can be defined as $\tau=v_0/|f^\mathrm{p}_j|$, with $f^\mathrm{p}_j$ being force density exerted on the bead by the fluid represented by the stress tensor $T_{ij}$, defined in Eq.~(\ref{eq:stress_tensor}):
 \begin{equation}\label{eq:numerical_forces}
    f^\mathrm{p}_j=\int_{\Gamma_\mathrm{p}} \mathrm{d}s\,\hat{n}_i T_{ij},
\end{equation}
with $\hat{n}_i$ being a versor normal to the bead edge $\Gamma_\mathrm{p}$. The force can be decomposed into $\bm{f}^\mathrm{p}=[f^d,f^l]$, i.e., drag and lift components, respectively. Having drag and lift forces calculated for a given numerical solution, via. Eq.~(\ref{eq:numerical_forces}), we can estimate the drag $M_d$ and lift $M_l$ coefficients via the equation $v^b_i=\frac{1}{\eta_s}(M_d\delta_{ij}-M_l\epsilon_{ij})f^\mathrm{p}_j$.

Secondly, in the FEM formulation we do not linearize the mass conservation equation~\ref{eq:cons_mass}, therefore, the term $\delta\rho\partial_k{}v_k$, that should not vanish for compressible fluid, is naturally present. Without this, it would not be possible to observe effects related to compressibility, such as odd viscosity~\cite{Ganeshan2017}. However, during the linearization this term is not preserved (cf. Eq.~(\ref{eq:cons_mass_lin})) and thus mass exchange process $\frac{\delta\rho}{\kappa}$ needs to be added. 
To reconcile numerical formulation and linearization, we estimate the mass exchange equivalent as $\kappa=\langle\partial_k{}v_k\rangle^{-1}$, with the average $\langle.\rangle$ integrated numerically in the area close to the bead. 

Lastly, to make numerical calculations a bit simpler, we assume the linearized form of the term $\epsilon_{ij}T_{ij}$ in the angular momentum conservation equation~(\ref{eq:cons_ang_momentum}). Finally, the system that we solve using the FEM is as follows:
\begin{align}
    \partial_t \rho + \partial_k (\rho v_k) &= 0,\nonumber\\
    \rho ( \partial_t + v_k \partial_k ) v_j &= \partial_i T_{ij},\nonumber\\
    \rho I ( \partial_t + v_k \partial_k ) \xi &= \partial_i C_i + 2\mu_r\!\left(\epsilon_{ij}\partial_i{}v_j-2\xi\right) - \frac{\rho I \xi}{\alpha},
    \label{eq:fem_system}
\end{align}
together with the EoS and the stress tensors $T_{ij}$, $C_i$, defined as in Eqs.~(\ref{eq:eo_state}-\ref{eq:couple_stress_tensor}), respectively. 

After setting up the finite domain formulation of the lift force problem and discussing the numerical method (FEM) used, let us analyze example results presented in Fig.~\ref{fig:fem_example}(c-f).
Subsequent columns present the velocity, pressure, density and microrotation fields. Some of the parameters used are listed on the right-hand side of the plots (unless otherwise specified, the rest of them have the following values: $\bar{\eta}_b=1$, $\bar{\mu}_r=0.4$, $\bar{I}=0.1$, $\bar{c}_1=2$, $\bar{\alpha}=0.5$) -- they can be used to control various regimes of fluid behavior. Fig.~\ref{fig:fem_example}(a) shows a typical solution for compressible fluid ($\bar{\chi}=4$) with an odd viscosity present ($\bar{\eta}_o=0.5$) and non-negligible coupling with the microrotation field ($\bar{\mu}_r=0.4$). Compressibility gives a characteristic increase (decrease) in fluid density $\rho$ in the area just in front (behind) of the bead. The density distribution is closely related to the pressure field $p$ (through the EoS) which is clearly visible on the plots. The microrotation field $\xi$ has a characteristic dipolar distribution with increasing/decreasing values at the front of the bead. Suppose we now switch off ($\bar{\eta}_o=0$) the odd viscosity term as in Fig.~\ref{fig:fem_example}(d), or change its sign ($\bar{\eta}_o=-0.5$) -- Fig.~\ref{fig:fem_example}(e), then the pressure (and also density) field solutions will (d) get symmetrized, or (e) will be mirror-symmetric (with respect to the center horizontal line). This is expected behavior of the antisymmetric odd term present in the system. At the same time, the microrotation field will get mirror-antisymmetrized -- see Figs.~\ref{fig:fem_example}(c) vs. (e). Moreover, if we now lift the compressibility condition, by putting large $\chi=10^4$ as in Figs.~\ref{fig:fem_example}(f), the pressure $p$ distribution remains similar, but now the density $\rho$ is homogeneous and the microrotation field $\xi$ takes the well-known distribution form (cf. Fig.~2 in~\cite{Hayakawa2000}). In case of no obstacle (bead) present, this will lead to the standard Poiseuille solution for the microrpolar fluid~\cite{lukaszewicz_micropolar_1999}. 

\begin{figure}
    \centering
    \includegraphics[width=0.95\linewidth]{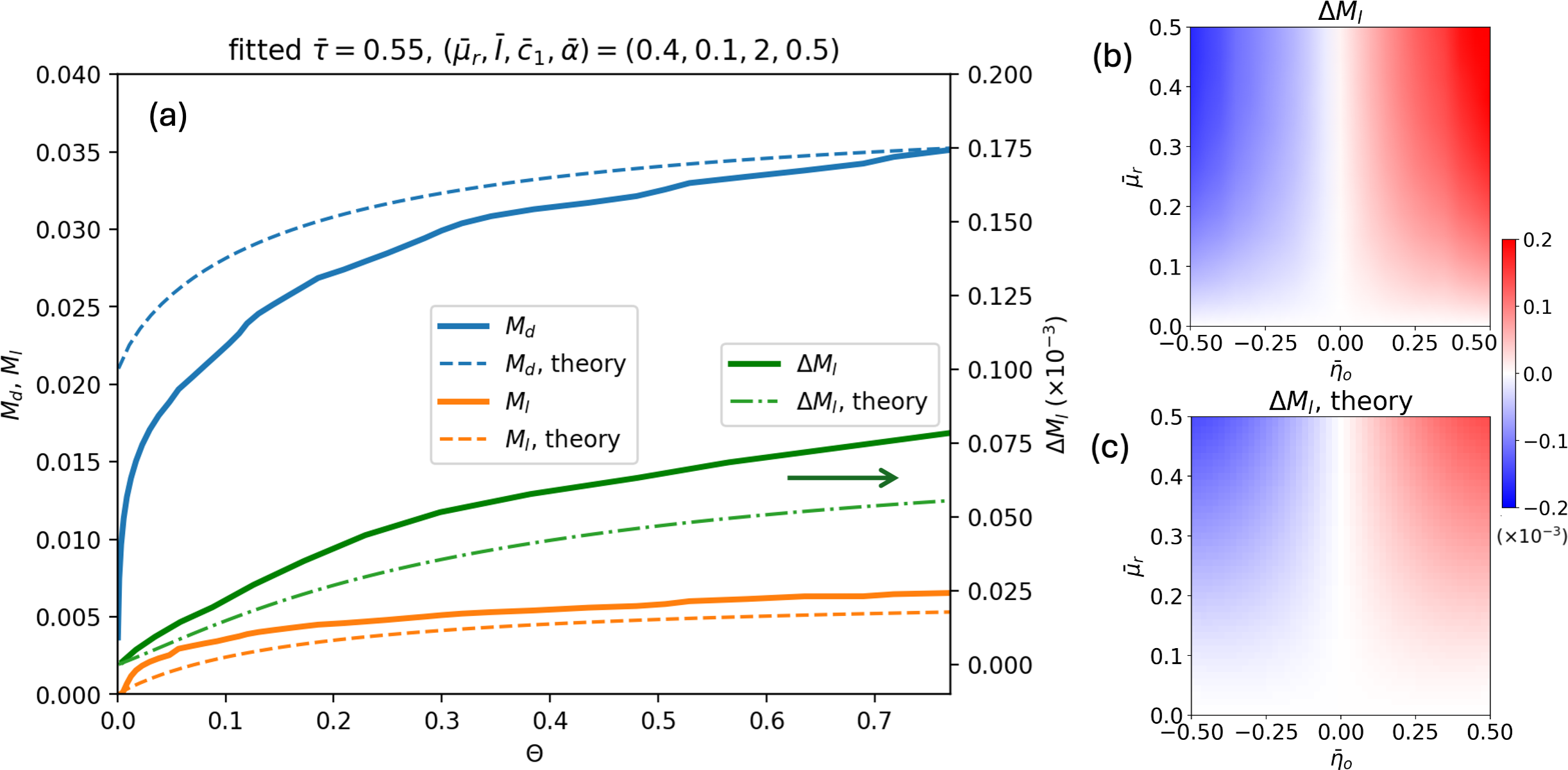}
    \caption{Comparison of forces calculated using the finite element method and obtained via the shell localization. (a) Drag $M_d$, lift $M_l$ force coefficients, and corrections to lift $\Delta{}M_l$ force coefficients due to coupling with the micropolar field are presented. Lift correction as a function of odd $\bar{\eta}_o$ and microrotation $\bar{\mu}_r$ couplings obtained in FEM (b) are compared with the shell localization analytical results (c).}
    \label{fig:fem_forces}
\end{figure}
Now we are ready to discuss the drag and lift force coefficients. Fig.~\ref{fig:fem_forces} shows drag $M_d$ (blue curves), lift $M_l$ (orange) force coefficients, and corrections to lift $\Delta{}M_l$ (green) force coefficient due to coupling with the micropolar field. Analytical results from the shell localization method (using Eqs.~(\ref{eq:drag_coeff}) and~(\ref{eq:lift_coeff})) are shown as dashed curves, while the FEM results are depicted by solid lines. The assumed parameters are the same as those for the calculations in Fig.~\ref{fig:fem_example}.
The calculated coefficients exhibit the expected behavior in the incompressible fluid limit, with  $M_d$ approaching a finite value and $M_l$ vanishing for $\Theta\rightarrow{}0$. The finite domain results qualitatively agree with the shell localization calculations in this $\Theta$ range. The correction $\Delta{}M_l$ term in the FEM case is slightly larger than the analytical counterpart, yet it correctly approaches zero as $\Theta$ becomes small.
If we now look at a map showing $\Delta{}M_l$ as a function of $(\bar{\eta}_o,\bar{\mu}_r)$ in Fig.~\ref{fig:fem_forces}(b,c) we observe that the lift correction increases with $\bar{\mu}_r$ as expected, however, it also changes sign along with $\bar{\eta}_o$ which also results in vanishing microrotation-induced lift correction with the odd term being zero. Both, FEM (b) and shell localization method (c) agree quite well. 

\section{Conclusions}

In this work, we have shown that compressible chiral granular materials are ideal for measuring lift forces due to odd viscosity. As such, the resulting experimental setups complement and extend previous proposals in Newtonian fluids with odd viscosity. Additionally, we have computed corrections from the microrotational viscosities, fully accounting for antisymmetric, gapped degrees of freedom in the micropolar medium.

To align the theoretical analysis with experimental conditions, we developed a finite element method that accommodates finite domain flows and compressibility in exact manner. Numerical results corroborate the approximate analytical considerations that bead tracers in an odd granular medium experience transverse forces, resulting from the underlying parity breaking of the medium.

Our analysis demonstrates that passive, chiral, compressible granular matter, when described by micropolar fluid dynamics, exerts a transverse force on a bead immersed in it, relative to the bead's direction of motion. This effect arises due to 'odd viscosity' present in the medium, a phenomenon linked to parity breaking. Such breaking, in turn, is caused by the chiral nature of the constituents within the medium. Importantly, the considerable size of these constituents in micropolar media means that this odd viscosity emerges without the need for activity. Moreover, active chiral granular media are expected to exhibit similar phenomena, analogous to behaviors observed in active Newtonian fluids.

\section*{Acknowledgements}
MB acknowledges the support of the Shanghai Municipal Science and Technology Major Project (Grant No. 2019SHZDZX01) and the support of the sponsorship from the Yangyang Development Fund. JP acknowledges support from Polish National Science Centre (NCN), under grant no. 2021/43/D/ST3/01989. PS was supported in part by the Polish National Science Centre (NCN) Sonata Bis grant 2019/34/E/ST3/00405, and the Unite! University alliance.
JZ acknowledges the support of the NSFC (No. 11974238 and No. 12274291) and the Shanghai Municipal Education Commission Innovation Program under No. 2021-01-07-00-02-E00138. JZ also acknowledges the support from the Shanghai Jiao Tong University Student Innovation Center.

\section*{Appendix}
\subsection*{FEM weak formulation}
To implement the FEM calculations we have used FEniCS library~\cite{Alnaes2015,Logg2010} which enables convenient expression of equations in their weak formulation through the UFL language~\cite{Alnaes2014}. Meshes were created using the Gmsh library~\cite{Geuzaine2009}. The weak formulation for the system~(\ref{eq:fem_system}) is as follows.

Let us start with a step for calculating the tentative velocity $\tilde{v}^\ast$:
\begin{align}
&\frac{1}{\Delta t}\left\langle\tilde{v}_j^\ast-\tilde{v}_j^n\middle| u_j \right\rangle+
\left\langle \tilde{v}_k^n \partial_k\!\left(\tilde{v}^n_j/\rho^n\right)\middle|u_j\right\rangle+\nonumber\\
&\left\langle T_{ij}\!\left(\tilde{v}^{n+\frac{1}{2}}_j,P^n,\xi^n\right)\middle|\varepsilon(u_j) \right\rangle-
\left\langle n_i T_{ij}\!\left(\tilde{v}^{n+\frac{1}{2}}_j,P^n,\xi^n\right)\middle| u_j \right\rangle_{\partial\Omega}=0\nonumber,\\
&\tilde{v}^{n+\frac{1}{2}}_j=\frac{\tilde{v}_j^\ast+\tilde{v}_j^n}{2\rho^n},
\end{align}
where $\varepsilon(u_j)=\frac{1}{2}(\partial_i u_j+\partial_j u_i)$ is the strain rate tensor, and $n_i$ is a versor normal to the computational box $\Omega$ boundary $\partial\Omega$.
In the above formula, we used the short-hand notation for inner products: $\left\langle u \middle| w \right\rangle=\int_\Omega \mathrm{d}^2x\,uw$, and $\left\langle u \middle| w \right\rangle_{\partial\Omega}=\int_{\partial\Omega} \mathrm{d}s\,uw$.
Replacing the test function $u_j$ with basis functions localized on the finite element mesh results in a discretized matrix form of the equation.
Note that to improve the numerical stability we rephrase the velocity field as $\tilde{v}_j=\rho v_j$.
Then we proceed with the pressure correction step, obtaining the updated value $P^{n+1}$:
\begin{align}
\left\langle\partial_k P^{n+1}\middle| \partial_k Q \right\rangle=
\left\langle\partial_k P^n\middle| \partial_k Q \right\rangle
-\frac{1}{\Delta t}\left\langle\partial_k\tilde{v}_k^\ast\middle|Q\right\rangle,
\end{align}
with $Q$ being a scalar-valued test function from the pressure space.
Now we are ready to perform the velocity correction step, resulting in the updated $\tilde{v}^{n+1}_j$:
\begin{align}
\left\langle \tilde{v}_j^{n+1}\middle| u_j \right\rangle=
\left\langle \tilde{v}_j^\ast\middle| u_j \right\rangle-
\Delta t\left\langle\partial_j(P^{n+1}-P^n)\middle| u_j \right\rangle.
\end{align}
The density is corrected in two sub-steps. In the first one we utilize the continuity equation:
\begin{align}
\left\langle\rho^\ast\middle| s \right\rangle=
\left\langle\rho^n\middle| s \right\rangle-
\Delta t\left\langle\partial_k\tilde{v}^{n+1}_k\middle|s \right\rangle,
\label{eq:numrho1}
\end{align}
where $s$ is a scalar test function from the density space. Then, we combine the previous sub-step, Eq.~(\ref{eq:numrho1}), giving the tentative density $\rho^\ast$, with the EoS:
\begin{align}
\rho^{n+1}=\rho^\ast w_\rho+\rho_0\left(\frac{1}{\chi}(P^{n+1}-P_0)+1\right)(1-w_\rho).
\label{eq:numrho2}
\end{align}
Note, that the above equation~(\ref{eq:numrho2}) is just an explicit formula for the updated density $\rho^{n+1}$.
The update weight parameter $w_\rho=0.9$ was tuned to stabilize the numerical solutions in an ``artificial'' time $\Delta t$ -- we are looking for steady-state solutions. 
Finally, the microrotation field update step is:
\begin{align}
&\frac{I}{\Delta t}\left\langle (\xi^\ast-\xi^n)\rho^n\middle|z\right\rangle+
I\left\langle\tilde{v}^n_k\partial_k\xi^n\middle|z\right\rangle+
c_1\!\left\langle\partial_k\xi^n\middle|\partial_k z\right\rangle+\nonumber\\
&\frac{I}{\alpha}\left\langle\xi^n\rho_n\middle| z\right\rangle-
2\mu_r\left\langle\epsilon_{ij}\partial_i\!\left(\tilde{v}^n_j/\rho^n\right)\middle| z\right\rangle+
4\mu_r\left\langle\xi^n\middle|z\right\rangle=0.
\label{eq:numxi}
\end{align}
In the Eq.~(\ref{eq:numxi}) we assume vanishing microrotation, i.e. $\xi=0$, on the computational box boundary $\partial\Omega$ (and the same no-slip condition at the bead edge). To ensure that the chosen conditions have no significant impact, we verified that changing the condition to an arbitrary value of $\xi=-5$ at the computational box boundary $\partial\Omega$ does not alter the microrotation distribution details near the bead disk (within a 1\% margin).
The tentative microrotation $\xi^\ast$ enters a formula for the updated microrotation $\xi^{n+1}$:
\begin{align}
\xi^{n+1}=\xi^\ast w_\xi+\xi^n (1-w_\xi),
\end{align}
with much slower update weight $w_\xi=0.1$. This closes the system of equations.

The described iterative process cycles through these five steps multiple times until convergence among all fields is reached.

\subsection*{Odd Newtonian fluid (with no micropolarity)}
\begin{figure}
    \centering
    \includegraphics[width=0.8\linewidth]{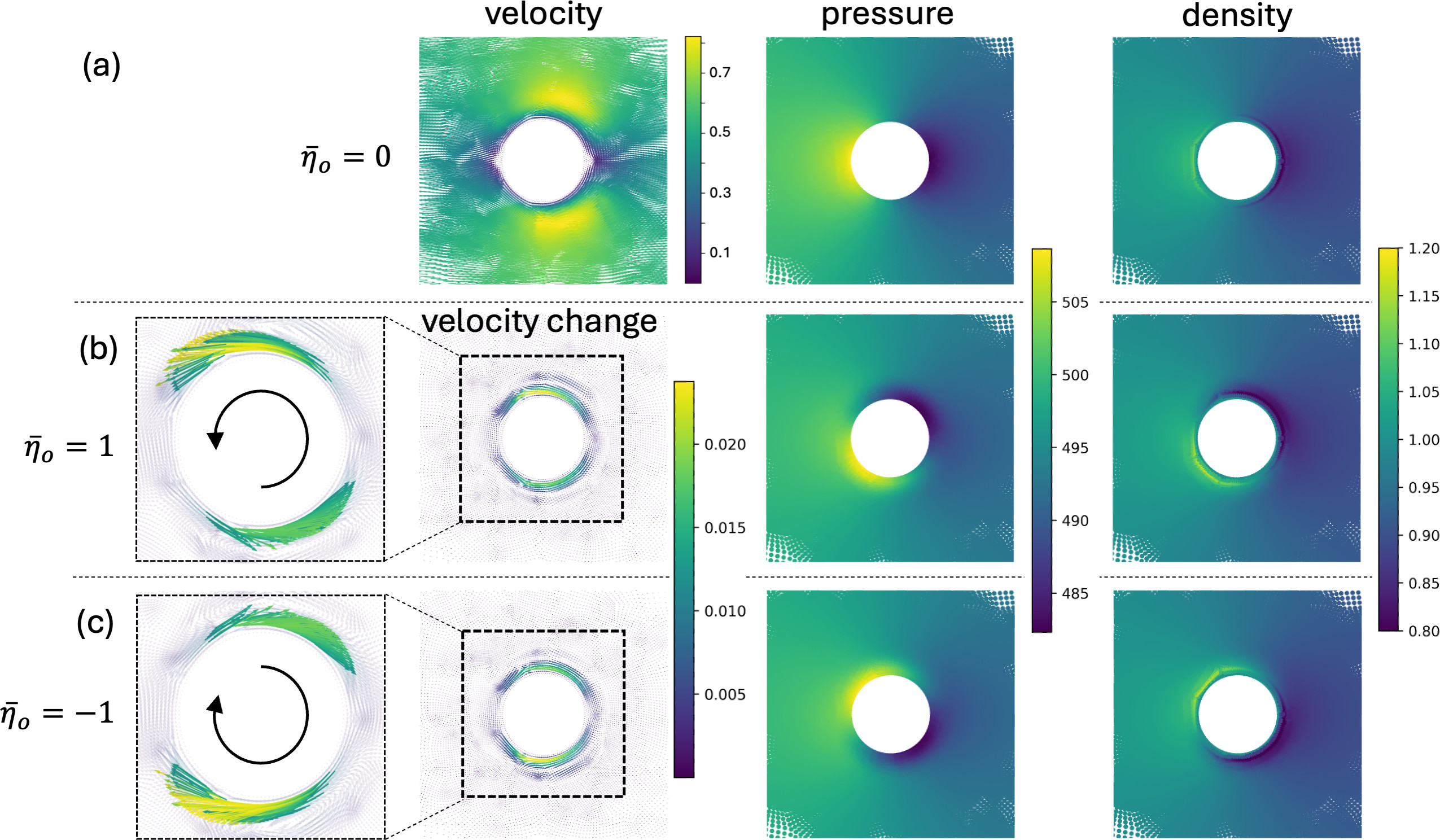}
    \caption{Finite element method results for Newtonian compressible fluid (without microrotation degree) flowing through a bead disk.
    Example solution (velocity, pressure, and density) fields are displayed for fluids with (a) zero oddity, (b) positive oddity, and (c) negative oddity. In panels (b,c) instead of the velocity field, we show the difference between the current velocity field and the field in case (a), i.e. without odd viscosity.}
    \label{fig:fem_justodd}
\end{figure}
For comparison, in Fig.~\ref{fig:fem_justodd} we additionally present results for Newtonian fluid only, i.e. not coupled with any microrotational degree of freedom. In Fig.~\ref{fig:fem_justodd}(a) we start with an example flow for the fluid without the odd viscosity term ($\bar{\eta}_0=0$). Then, we can observe that the addition of the odd viscosity ($\bar{\eta}_0=1$) in Fig.~\ref{fig:fem_justodd}(b) results in the emergence of the flow velocity components that would force the bead disk to rotate. Finally, changing the sign of the odd viscosity term ($\bar{\eta}_0=-1$) in Fig.~\ref{fig:fem_justodd}(c) reverses the direction of the vortex. The following parameters were adopted for the simulations in Fig.~\ref{fig:fem_justodd}: $\bar{\eta}_b=1$, $\bar{\chi}=4$.

\bibliography{main}

\end{document}